\documentclass[twocolumn,prl]{revtex4-2}

\usepackage{graphicx}
\usepackage{dcolumn}
\usepackage{bm}
\usepackage{multirow}
\usepackage{color}
\usepackage[normalem]{ulem}
\usepackage{amsmath}
\usepackage{gensymb}
\usepackage[colorlinks=true]{hyperref}

\renewcommand\thefigure{\arabic{figure}}

\begin{document}

\title{Raman Polarization Switching in CrSBr}
\author{Priyanka Mondal$^{1}$}
\author{Daria I. Markina$^{1}$}
\author{Lennard Hopf$^{1}$}
\author{Lukas Krelle$^{1}$}
\author{Sai Shradha$^{1}$}
\author{Julian Klein$^{2}$}
\author{Mikhail M. Glazov$^{3}$}
\author{Iann Gerber$^4$}
\author{Kevin Hagmann$^{1}$}
\author{Regine v. Klitzing$^{1}$}
\author{Kseniia Mosina$^{5}$}
\author{Zdenek Sofer$^{5}$}
\author{Bernhard Urbaszek$^1$}
\email{bernhard.urbaszek@pkm.tu-darmstadt.de}

\affiliation{\small$^1$Institute for Condensed Matter Physics, TU Darmstadt, Hochschulstraße 6-8, D-64289 Darmstadt, Germany
}
\affiliation{\small$^2$Department of Materials Science and Engineering, Massachusetts Institute of Technology, Cambridge, Massachusetts 02139, United States}
\affiliation{\small$^3$Ioffe Institute, 194021 St. Petersburg, Russia
}
\affiliation{\small$^4$Universit\'e de Toulouse, INSA-CNRS-UPS, LPCNO, 135 Avenue Rangueil, 31077 Toulouse, France}
\affiliation{\small$^5$Department of Inorganic Chemistry, University of Chemistry and Technology Prague, Technicka 5, 166 28 Prague 6, Czech Republic}

\begin{abstract}

Semiconducting CrSBr is a layered A-type antiferromagnet, with individual layers antiferromagnetically coupled along the stacking direction. Due to its unique orthorhombic crystal structure, CrSBr exhibits highly anisotropic mechanical and optoelectronic properties acting itself as a quasi-1D material. CrSBr demonstrates complex coupling phenomena involving phonons, excitons, magnons, and polaritons. Here we show through polarization-resolved resonant Raman scattering the intricate interaction between the vibrational and electronic properties of CrSBr. For samples spanning from few-layer to bulk thickness, we observe that the polarization of the A$_g^2$ Raman mode can be rotated by 90 degrees, shifting from alignment with the crystallographic $a$ (intermediate magnetic) axis to the $b$ (easy magnetic) axis, depending on the excitation energy. In contrast, the A$_g^1$ and A$_g^3$ modes consistently remain polarized along the $b$ axis, regardless of the laser energy used.  We access real and imaginary parts of the Raman tensor in our analysis, uncovering resonant electron-phonon coupling. 
\end{abstract}

\maketitle

\begin{figure*}
\includegraphics[width=0.85\linewidth]{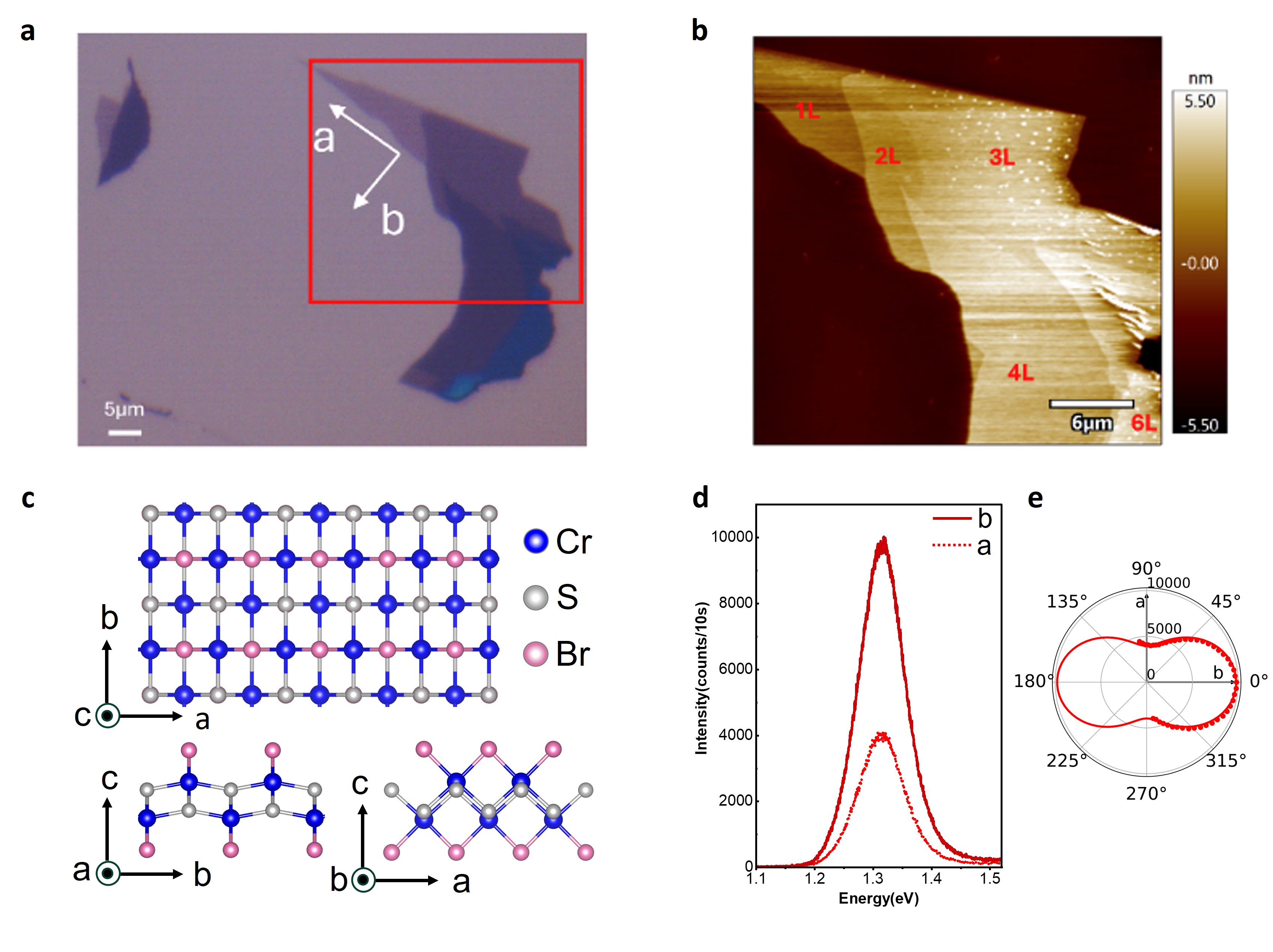}
\caption{\label{fig:fig1} \textbf{CrSBr crystal structure and photoluminescence.} (a) Optical image of investigated CrSBr sample on SiO$_2$/Si substrate. The white arrows indicate the $a$ and $b$ crystallographic axes of the depicted flake, and the red box denotes the area covered by the AFM scan. (b) Atomic force microscopy image of the sample. The atomic layer thicknesses of different areas are marked in red. (c) Sketch of CrSBr crystal structure, crystallographic axes $a$, $b$, and $c$ are indicated. Red arrows in the top panel show an in-plane spin orientation in a single layer. (d) Photoluminescence spectra of four-layer (4L) thick CrSBr flake at T = 300 K, PL emission is polarized along the $a$ (dashed line) and $b$ (solid line) crystalographic directions. (e) Polar plot of the  PL intensities as a function of light polarization orientation to a $b$ axis (0 degrees).}
\end{figure*}

\textbf{Introduction.---} Investigating new air-stable materials that intrinsically exhibit both semiconducting behavior and magnetic ordering is imperative to advancing electron spin-based technologies \cite{telford2020layered,huang2018electrical,gibertini2019magnetic,dyakonov2017basics,ziebel2024crsbr}. Recent progress in such materials mainly concern van der Waals materials such as  FePS$_3$ \cite{wang2016raman}, Cr$_2$Ge$_2$Te$_6$ \cite{gong2017discovery, tian2016magneto}, CrI$_3$ \cite{huang2017layer, lado2017origin}, the latter of which is not air stable. Among the most promising materials of this family is chromium sulfide bromide (CrSBr). CrSBr is an A-type antiferromagnet with a Neel temperature of 130 K with tunable interlayer coupling \cite{dirnberger2023magneto,wilson2021interlayer,klein2023bulk,ruta2023hyperbolic,telford2020layered,tabataba2024doping}. It is a two-dimensional layered material with orthorhombic crystal structure of $Pmmn$ space group and D$_{2h}$ point group~\cite{torres2023probing}. A single layer of CrSBr is built up by two planes of chromium and sulfur atoms sandwiched between layers composed of only bromine atoms (Fig. ~\ref{fig:fig1}c). Due to its pronounced orthorhombic crystal structure and relatively weak interlayer hybridization, the material's physical properties are expected to exhibit a quasi-1D character. Current research aims to explore this pronounced anisotropy manifested, among other factors, in a 50-fold difference in effective carrier mass between the $a$ and $b$ axis directions \cite{klein2023bulk}. 

The exact interplay between anisotropic electronic band structure effects and excitons in optical transitions across the bandgap is not yet well understood and is currently actively investigated \cite{klein2023bulk,PhysRevResearch.5.033143,bianchi2023paramagnetic,PhysRevB.108.195410,watson2024giant,smolenski2024large,komar2024colossal,cenker2022reversible,doi:10.1021/acs.jpclett.4c00968,doi:10.1021/acs.nanolett.3c05010}.  Exciton states are sensitive to magnetic order and vice versa, making this a key area of research for understanding the properties and applications of CrSBr. A powerful technique to probe the symmetry of layered materials is Raman spectroscopy \cite{ferrari2006raman,zhang2015phonon,pimenta2021polarized}. In our experiments, we investigate the coupling between the vibrational modes and electronic properties of CrSBr. In polarization-resolved Raman spectroscopy, we uncover strong variations of the polarization angle (with respect to the magnetic easy axis) and Raman scattering intensity for the A$_g^2$ mode as we vary the laser energy from 1.58 to 2.33 eV. Surprisingly this behaviour is observed for samples from 4 monolayers to bulk thickness, indicating that the relevant electronic states do not evolve considerably as a function of the number of layers. By extracting the real and imaginary elements of the Raman tensor we quantify the anisotropy and electron-phonon coupling in the crystal. \\

\begin{figure*}
\includegraphics[width=1\linewidth]{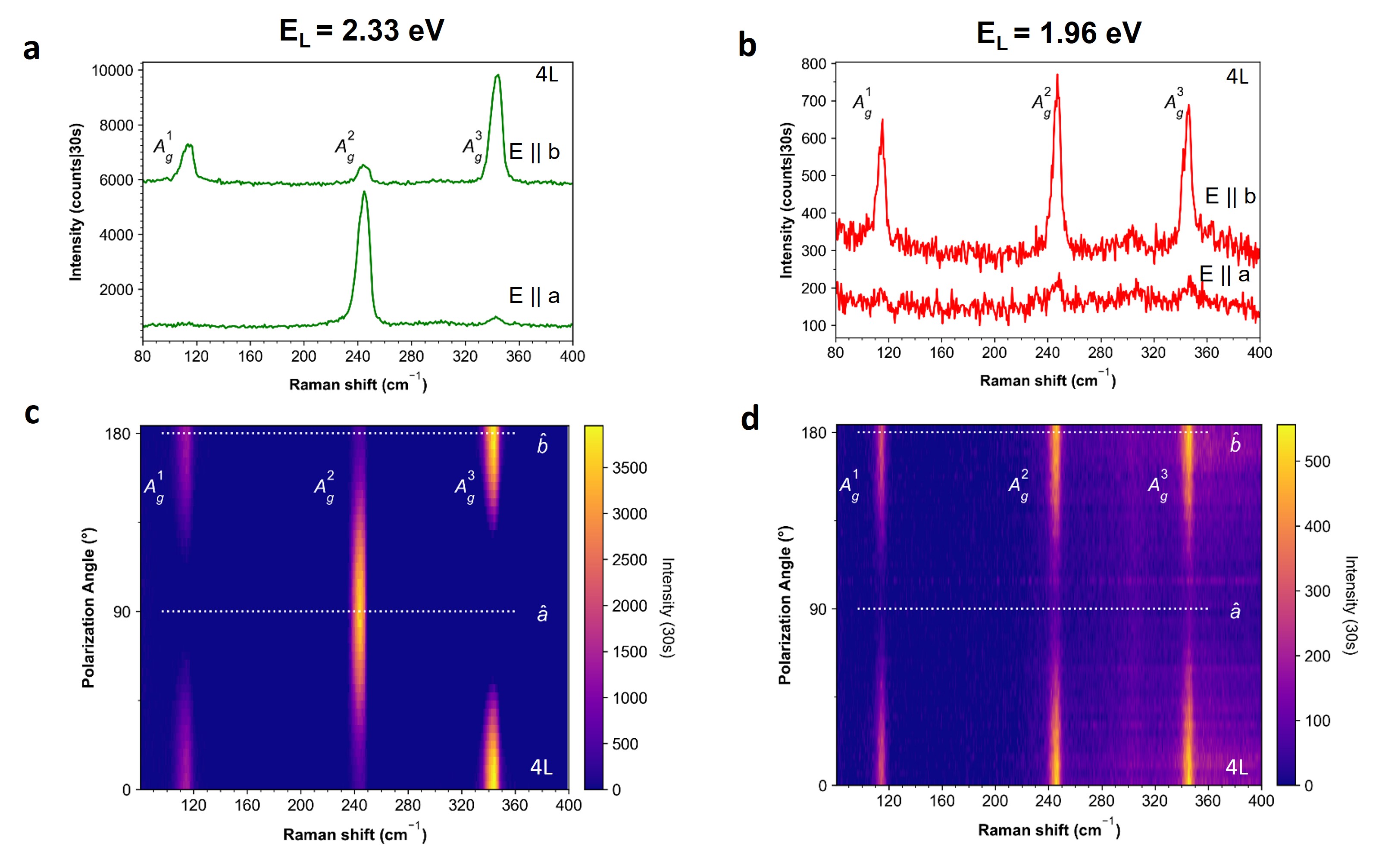}
\caption{\label{fig:fig2} \textbf{Raman modes polarization switching at different laser energies.} All results are obtained from a four-monolayer (4L) thick CrSBr sample. (a, b) Raman spectra at T = 300 K under an excitation laser energy of (a) $E_L$ = 2.33 eV and (b) $E_L$ = 1.96 eV for polarization along the $b$ axis (top) and $a$ axis (bottom). (c, d) Color mapping of Raman scattering signal for (c) $E_L$ = 2.33 eV and (d) $E_L$ = 1.96 eV as a function of polarization orientation angle, $a$ and $b$ axes are indicated as horizontal dotted lines and correspond to the individual spectra shown in (a, b). Under excitation of 2.33 eV A$_g^2$ mode reveals the polarization along $a$ direction rotated by 90 degrees concerning A$_g^1$ and A$_g^3$ modes polarized along $b$ axis. While under 1.96 eV all phonon modes are polarized along the $b$ axis.}
\end{figure*}

\textbf{Experimental Results.---}
As a van der Waals material CrSBr can be exfoliated to monolayer-thick flakes \cite{huang2015reliable}. Typically exfoliated CrSBr flakes have a rectangular shape with the long edge oriented along the $a$ axis and the short edge aligned along $b$, its (magnetic easy) axis. 
Fig.~\ref{fig:fig1}a shows an exfoliated CrSBr flake on an 85 nm SiO$_2$/Si substrate, the white arrows indicate the crystal axes orientation. The thickness of exfoliated samples was estimated using atomic force microscopy (AFM, see methods), as shown in Fig.~\ref{fig:fig1}b and Fig.~\ref{fig:figS1}.\\
\indent Polarization-resolved optical spectroscopy on exfoliated flakes is employed to probe the anisotropic optical transitions in CrSBr \cite{pimenta2021polarized,shree2021guide}. The room temperature linear polarization-resolved photoluminescence (PL) spectra of a four-monolayer (4L) thick CrSBr sample are represented in Fig.~\ref{fig:fig1}d. At room temperature, PL emission is centered at 1.31~eV with FWHM = 90 meV. We observe stronger PL emission along the $b$ direction than along the $a$ direction (Fig.~\ref{fig:fig1}e). This observation remains valid across the investigated excitation laser energy range ($E_L$ = 1.96~eV to 2.33~eV, Fig.~\ref{fig:figS2}) and sample thicknesses, from a few layers to bulk. Polarization-resolved PL spectroscopy enables the determination of the $a$ and $b$ axis orientation, providing a reference for subsequent Raman spectroscopy analysis. \\
\indent In total, there are 18 phonon modes of CrSBr in the first Brillouin zone: 15 optical modes, and 3 acoustic ones. Based on the symmetry analysis of the Raman tensors, only the three optical A$_g$ modes are expected to be observed in first-order Raman scattering \cite{torres2023probing,pimenta2021polarized} and our experimental investigation is focused on these three modes. 
The A$_g$ modes correspond to out-of-plane lattice vibrations of chromium, sulfur, and bromine atoms, with varying interlayer and intralayer character \cite{torres2023probing}.\\
\indent In our Raman scattering experiments, the linearly polarized laser light is incident along the $c$ crystallographic axis on the sample, and the laser polarization axis is rotated to cover all angles in the crystallographic $ab$ plane, as shown in the schematic of the measurement setup in Fig.~\ref{fig:figS3}. Fig.~\ref{fig:fig2} shows the Raman scattering data on a 4L-thick flake at room temperature for different excitation energies (E$_L$ = 2.33 and 1.96 eV). Experimental results reveal well-separated out-of-plane modes A$_g^1$ (114~cm$^{-1}$), A$_g^2$ (244 ~cm$^{-1}$) and A$_g^3$ (344~cm$^{-1}$), which is consistent with the result shown in previous studies on few-layer CrSBr \cite{klein2023bulk,pawbake2023raman}. In the top spectrum (Fig.~\ref{fig:fig2}a,b) excitation polarization is aligned along the $b$ axis of CrSBr, in the bottom spectrum it is aligned along the $a$ axis. 
For excitation with the laser energy of 2.33~eV, which exceeds the single-particle band gap, the A$_g^1$ and A$_g^3$ modes show maximum signal along the $b$ axis, and the A$_g^2$ mode along the $a$ axis (Fig.~\ref{fig:fig2}c), which agrees with previous reports using E$_L=2.33$~eV \cite{klein2023bulk}. In contrast, for excitation at E$_L$ = 1.96 eV, all three primary modes (A$_g^1$, A$_g^2$, A$_g^3$) exhibit maximum intensity along the $b$ axis, and the Raman scattering signal for these modes essentially vanishes for incident polarization along the $a$ axis, as can be seen in Fig.~\ref{fig:fig2}d. The Raman active modes A$_g^1$ and A$_g^3$ are polarized along the $b$ axis for both excitation energies used in Fig.~\ref{fig:fig2}, whereas mode A$_g^2$ switches polarization by 90$^{\circ}$ with variation of the laser energy. \\
\begin{figure*}[t]
\includegraphics[width=1\linewidth]{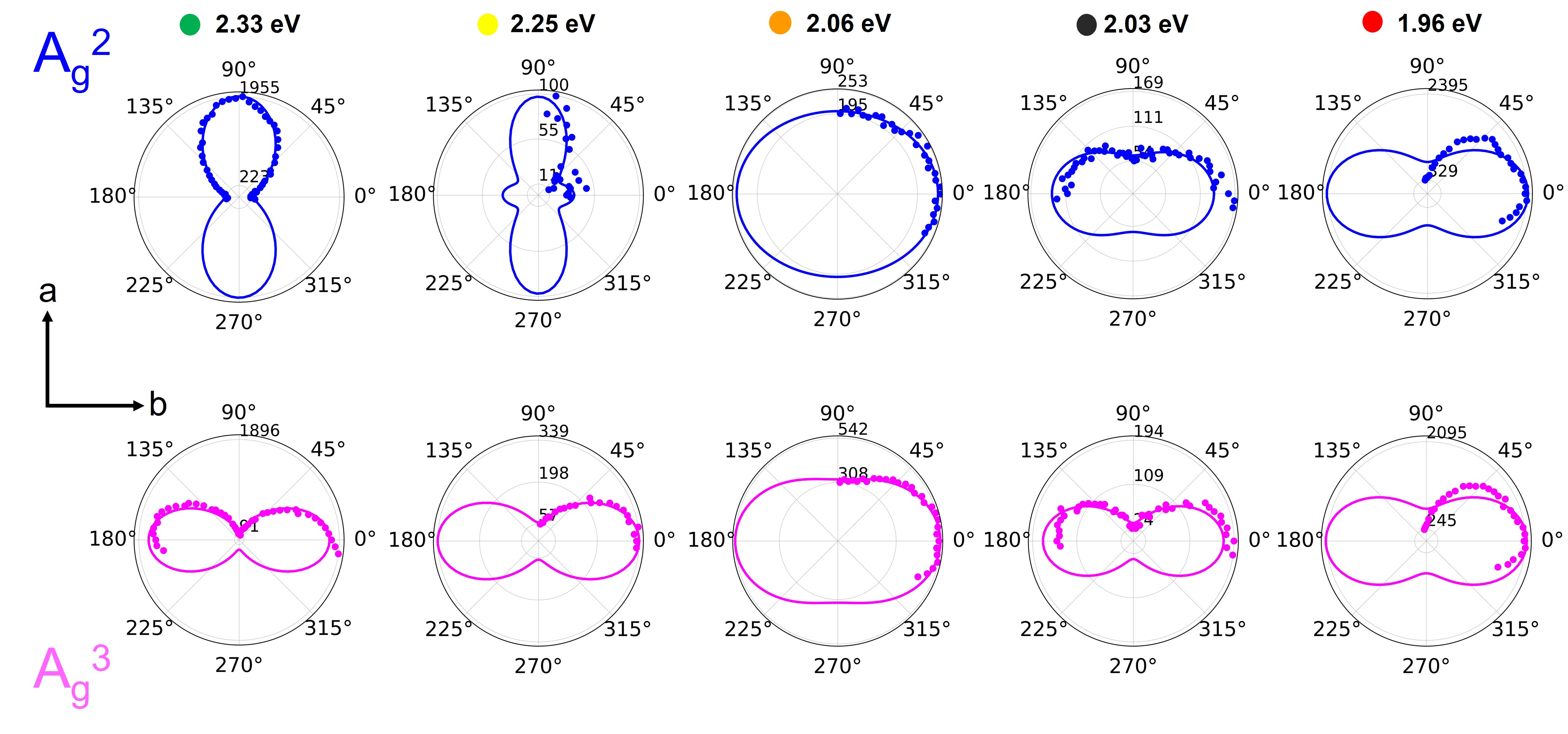}
\caption{\label{fig:fig3} \textbf{Polarization dependence of Raman modes of 6L CrSBr for different excitation energies.} The top panel presents the polar plot of Raman scattering signal intensity dependent on the polarisation rotation angle of A$_g^2$ (blue color), while the bottom panel shows the same for  A$_g^3$ mode (pink color) at different excitation energies ranging from 2.33 to 1.96 eV (each column reflects a certain excitation energy). Dots correspond to an experimental spectrum, and the solid lines are fitting curves using equation \eqref{fit}. The angle of the polarization orientation of the Raman signal is measured from the direction of the $b$ axis (0 degrees). The radial axis corresponds to the signal intensity, which differs for each plot, the center point is 0 counts on all plots. The A$_g^2$ mode polarization behavior switches from being polarized along $a$ to along $b$ direction at around 2.07 eV excitation energy.}
\end{figure*}
\indent To gain further insights into the polarization switching of the A$_g^2$ mode, we conducted polarized Raman measurements using various laser excitation energies between 1.96 eV and 2.33 eV. These measurements were carried out using a tunable laser (see Methods Section), which enables the extraction of polarization dependencies (Fig.~\ref{fig:figS4}). This experimental approach allows insights into electron-phonon interactions as shown for various anisotropic materials such as CrOCl~\cite{zhang2019magnetism}, black phosphorus~\cite{ling2016anisotropic,mao2016birefringence,ribeiro2015unusual,wu2015identifying,mao2019direct}, ReS$_2$~\cite{mccreary2017intricate}, ReSe$_2$~\cite{wolverson2014raman}, GaTe~\cite{huang2016plane}, PdSe$_2$~\cite{luo2022excitation} and also MoS$_2$~\cite{miller2019tuning,PhysRevB.87.115413}.\\
\indent Polarization dependences of the A$_g^2$ (Fig.~\ref{fig:fig3}, top panel) and A$_g^3$ modes (Fig.~\ref{fig:fig3}, bottom panel) in 6L-thick CrSBr for excitation energies of 2.33, 2.25, 2.06, 2.03, and 1.96~eV are presented. The angles are measured relative to the $b$ axis direction. Obtained data shows that the polarization of the A$_g^2$ mode gradually changes with excitation energy. The A$_g^2$ mode exhibits maximum intensity along the $a$ axis for 2.33~eV and 2.25~eV excitation energies. For a laser excitation of 2.06~eV, the A$_g^2$ polarization is drastically different: the scattered light along the $a$ and $b$ direction (and for orientations in-between) have almost equal intensities, resulting in a close to circular polar plot. We discuss this later in terms of the Raman tensor contributions along the $a$ and $b$ axis. For laser excitation energies below $\sim$2.06 eV, the polarization of A$_g^2$ mode shows maximum intensity along $b$ axis, changing its polarization by 90$^{\circ}$. It is important to mention, that at lower excitation energies (E$_L$ = 1.58 eV) polarization of the A$_g^2$ mode switches back along $a$ crystallographic axis as for energies higher than 2.06 eV ( Fig.~\ref{fig:figS5}, ~\ref{fig:figS7}). This excitation energy-dependent switching in Raman spectroscopy can be compared with the results of Photoluminescence Excitation Spectroscopy (PLE), showing resonance behavior at around 1.87 eV and out of resonance at E$_L$ = 2.33 eV and E$_L$ = 1.58 eV when exciting with light polarized along $b$ axis and more complex behavior under the excitation polarization along $a$ axis with two resonances at $\sim$1.65 and $\sim$1.95 eV (see Fig.~\ref{fig:figS8}). The drastic difference in the light-matter interaction along different crystallographic axes is reflected in the linear polarization degree \(P_{lin} = (I_b - I_a)/(I_b + I_a)\) (see Fig.~\ref{fig:figS8}). In contrast, the A$_g^3$ mode consistently exhibits the strongest intensity when the incident polarization is aligned with the CrSBr $b$ axis for all the excitation energies between 1.96 and 2.33 eV (see Fig.~\ref{fig:figS5},~\ref{fig:figS6},~\ref{fig:figS7} in Supplementary Information for additional excitation energies for both A$_g^2$ and A$_g^3$ modes). Thus, in CrSBr the polarization axis of the A$_g^2$ Raman mode can be switched between the $b$ and $a$ directions of the CrSBr crystallographic axes depending on the laser excitation energy.\\
\indent Besides the excitation energy, the thickness of the sample could also affect the Raman scattering results. To investigate it, we conduct the polarization-dependent Raman scattering measurements for samples of different thicknesses ranging from $\sim$3.5 to 62 nm (Fig.~\ref{fig:fig4}, top panel). All samples were excited with the energies of 2.33 eV (Fig.~\ref{fig:fig4}, middle panel) and 1.96 eV (Fig.~\ref{fig:fig4}, bottom panel). Analogous to the 6L sample (Fig.~\ref{fig:fig3}) laser excitation at 1.96 eV for all thicknesses results in the polarization of all three primary Raman modes along the $b$ axis, which corresponds to 0$^{\circ}$. In contrast, for an excitation with E$_L$ = 2.33 eV, both the A$_g^1$ and A$_g^3$ modes in all four samples shown in Fig.~\ref{fig:fig4} exhibit maximum intensity at 0 $^{\circ}$ along $b$, while the A$_g^2$ mode shows maximum intensity at 90$^{\circ}$ along $a$. We experimentally observe a consistent polarization switching of the A$_g^2$ Raman mode between the $a$ and $b$ crystallographic axes of CrSBr, depending on the laser excitation energy, across a wide range of CrSBr flake thicknesses. \\
\begin{figure*}
\includegraphics[width= 1\linewidth]{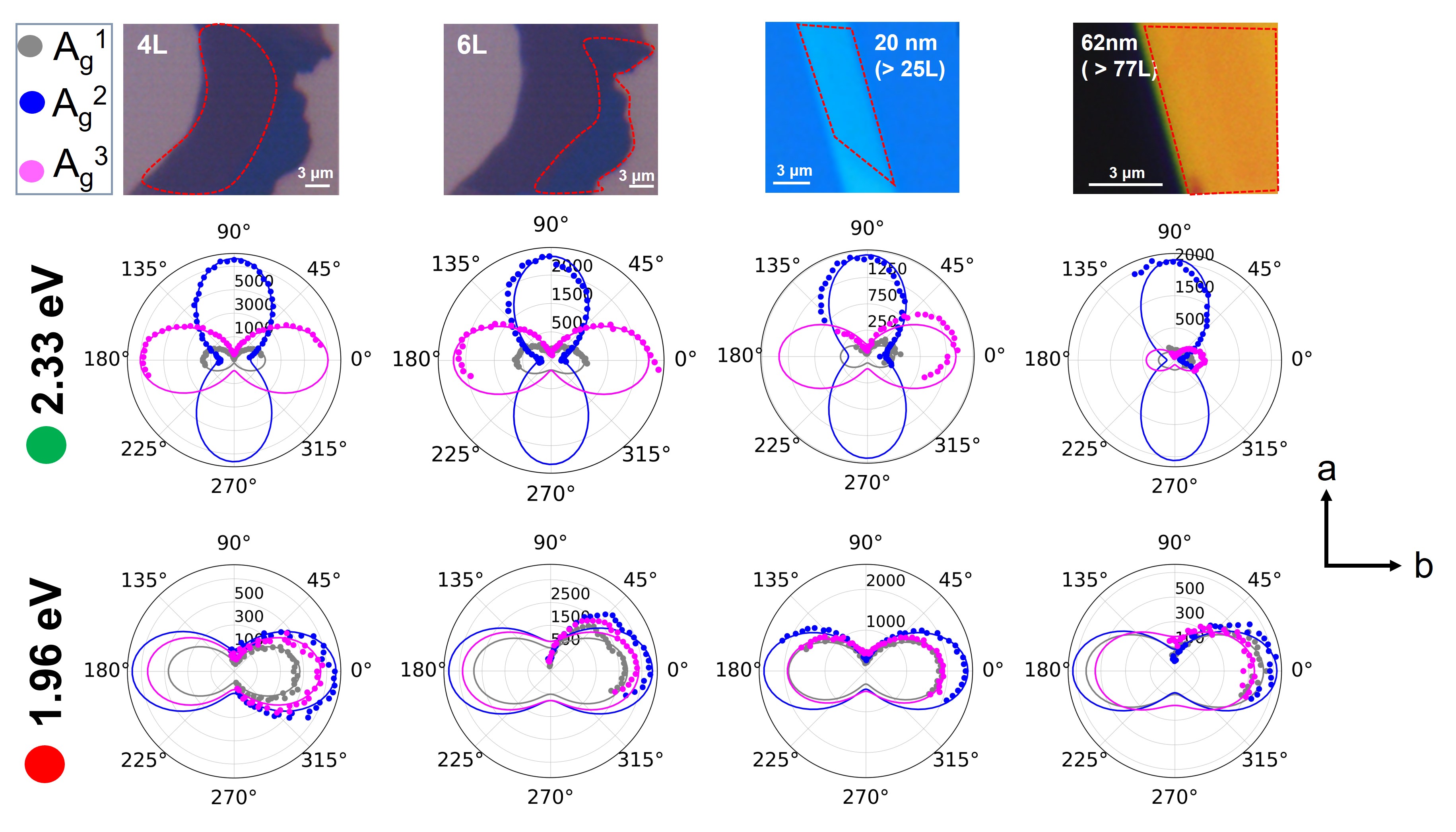}
\caption{\label{fig:fig4} \textbf{Polarization dependence of Raman modes for different sample thickness.} The top panel demonstrates bright field imaging of the samples of different thicknesses from $\sim$3.6 to 62 nm, columns under each image correspond to the results measured on the area marked with a red dashed line. The middle panel shows the polarization polar plots of all three Raman modes at the excitation of 2.33 eV, and the bottom panel illustrates the same Raman modes under 1.96 eV excitation. Dots correspond to experimental results, and the solid lines are fitting curves using equation \eqref{fit}. The angle of the polarization orientation of the Raman signal is measured from the direction of the $b$ axis (0 degrees). The radial axis corresponds to the signal intensity, which differs for each plot. The polarization orientation of each Raman mode is independent of the sample thickness.}
\end{figure*}
\indent In addition to the polarization state of all the modes mentioned above, analyzing the relative Raman scattering intensities gives additional insights. For E$_L$=1.96 eV, all modes possess comparable signal intensities (Fig.~\ref{fig:fig4}, bottom panel). This is not the case for 2.33 eV excitation. A$_g^2$ and A$_g^3$ modes demonstrate similar intensities for thicknesses from 3.5 to 20 nm (Fig.~\ref{fig:fig4}, middle panel). For thicker samples, optical interference effects start to play a role in the light-matter-interaction, which must be explicitly considered when evaluating the Raman tensor \cite{kim2015anomalous, ling2016anisotropic, luo2022excitation}. For a sample thickness of 62 nm, the A$_g^2$ Raman signal is considerably stronger than the A$_g^3$ one. \\

\textbf{Discussion.---} We demonstrated experimentally that the A$_g^2$ phonon mode is polarized differently from the A$_g^1$ and A$_g^3$ modes. The Raman tensor $\hat R(E_L)$ that describes the intensity of the scattered light $I_s \propto |\mathbf e_i \hat R \mathbf e_s|^2$ as function of polarization vectors of the incident, $\mathbf e_i$, and scattered, $\mathbf e_s$, fields and the excitation laser energy $E_L$ depends on both the crystal symmetry and electron-phonon coupling \cite{lin2024strong,peter2010fundamentals}. For $A_g$-symmetry modes the non-zero components of the Raman tensor are $a=R_{xx}$, $b=R_{yy}$, and $c=R_{zz}$ with $x$, $y$, and $z$ being the principal axes of the system ($a$-, $b$-, and $c$-axis, respectively). Phenomenologically, the response of $A^1_g$ and $A^3_g$ is dominated by $R_{yy}$ (along the $b$-axis). The response of $A^2_g$ shows sizeable contributions of both $R_{xx}$ and $R_{yy}$ and for certain incident laser energies $R_{xx}$ dominates the Raman process. The contribution of these two Raman tensor elements to the scattering intensity for different wavelengths in the range from 1.58 to 2.33 eV can be seen in  Fig.~\ref{fig:figS9}, ~\ref{fig:figS10}.
Microscopically, the Raman scattering in the optical range of incident laser frequencies is mediated by electronic excitations. The Raman intensity can be calculated by Fermi's golden rule transition probability for the third-order time-dependent perturbation process~\cite{pimenta2021polarized}. This gives the Raman tensor elements $R_{ij}^{\mu}$ for the phonon branch $\mu$ that depend on the laser excitation energy $E_L$ according to \cite{peter2010fundamentals}:
\begin{widetext}
\begin{equation}
    \label{Raman}
    R_{ij}^{\mu}(E_L)=\sum_{n,n'}\frac{\langle 0|\mathbf e_s^*\cdot \hat{\mathbf{p}}|n'\rangle\langle n'|H_{el-ph}^{\mu}|n\rangle \langle n|\mathbf e_i\cdot \hat{\mathbf{p}}|0\rangle}{[E_L-E_{n}+i\Gamma_n][E_L-E_{ph}^{\mu}-E_{n'}+i\Gamma_{n'}]}
\end{equation}
\end{widetext}
Here $|n\rangle$ and $|n'\rangle$ are the eigenstates of electronic excitations of the crystals, namely, electron-hole pairs, $E_{n}$, $E_{n'}$ are their energies, $\Gamma_n$, $\Gamma_{n'}$ are their damping. The interaction of crystal with light is described by the matrix elements of the momentum operator (e.g., $\langle n|\mathbf e_i\cdot \hat{\mathbf{p}}|0\rangle$) and $H_{el-ph}^{\mu}$ stands for the Hamiltonian of the electron interaction with the phonon mode $\mu$ with the energy $E_{ph}^\mu$.\\
For fitting of the polarisation-dependent Raman scattering intensities $I_{||}^{A_g}$ in  Fig.~\ref{fig:fig3},~\ref{fig:fig4} we used the following function for the parallel polarization configuration used in our experiment \cite{pimenta2021polarized}: 
\begin{equation}
    \label{fit}
    I_{||}^{A_g}({\theta})=(a{\cdot}\text{cos}^2{\theta}+b{\cdot}\text{cos}{\phi}_{ab}{\cdot}\text{sin}^2{\theta})^2+b^2{\cdot}\text{sin}^4{\theta}{\cdot}\text{sin}^2{\phi}_{ab}
\end{equation}
where \(a=R_{xx}\) and \(b=R_{yy}\) are non-zero components of Raman tensor, responsible for $a$- and $b$- crystallographic axis, $\theta$ is an angle between the incident light polarization and the $a$-axis, \(\phi_{ab}=(\phi_{b}-\phi_{a})\) is a relative phase between $b$ and $a$ components of A$_g$ tensor. Taking into account the imaginary part via $\phi$$_{ab}$ allows fitting all measurements including the four-fold shape for the A$_g^2$ mode for E$_L$ = 2.25 eV (Fig.~\ref{fig:fig3}). The requirement to consider the imaginary part of the Raman tensor element for fitting of A$_g$ modes is obvious from the comparison of two fitting approaches with and without $\phi$$_{ab}$ in Fig.~\ref{fig:figS5}, ~\ref{fig:figS6}, for additional numerical parameters from fitting see in Supplementary Information Fig.~\ref{fig:figS11}, ~\ref{fig:figS12}.\\
\indent Our measurements on polarization switching indicate that the A$_g^2$ mode couples differently to the electronic states compared to the A$_g^1$ and A$_g^3$ modes. Calculations (\cite{klein2023bulk, PhysRevResearch.5.033143}) suggest that excitonic states polarized preferentially along the $a$ axis exist at energies around 1.9 eV. This contrasts with the lowest-lying exciton states, which are polarized along the $b$ axis, as seen in the PL measurements in Fig.~\ref{fig:fig1}d. We can speculate that the $A_g^2$ mode is coupled stronger to these transitions. It is also consistent with the fact that the fitting of the Raman scattering polarization dependence in Fig.~\ref{fig:fig4} with \eqref{fit} requires to account for both real and imaginary parts of the Raman tensor components suggesting that a resonance or close-to-resonance condition occurs in Eq.~\eqref{Raman} where $|E_L - E_n|\lesssim \Gamma_{n}$ or $|E_L - E_{ph}^\mu - E_{n'}|\lesssim \Gamma$. The exact bandgap and energy position of excitonic states depend on layer thickness, however, the observed polarization switching of the A$_g^2$ mode is consistent across all thicknesses. Further resonant experiments at cryogenic temperatures would help target specific exciton states. It is important to note that the exact energy order and characteristics (dark versus bright states) of the excitonic states are still being investigated through various experimental and theoretical approaches in this complex system \cite{bianchi2023paramagnetic,PhysRevB.108.195410,watson2024giant,smolenski2024large,komar2024colossal,cenker2022reversible,doi:10.1021/acs.jpclett.4c00968,doi:10.1021/acs.nanolett.3c05010}.\\

\indent \textbf{Conclusion.---} The intrinsic anisotropy in the crystal structure of CrSBr impacts photoluminescence (PL) and Raman scattering polarization. We find that, across a range of samples with different thicknesses, the PL emission is oriented along the $b$ axis for various laser excitation energies. In contrast, in Raman spectroscopy, different excitation wavelengths reveal that the Raman-active modes A$_g^1$,  A$_g^2$, and  A$_g^3$ possess distinct polarization properties. This indicates an intricate interplay between electron-phonon interactions and crystal symmetry in this layered material. Our findings demonstrate that the  A$_g^2$  mode can switch its polarization axis between the $a$ and $b$ crystallographic directions as a function of excitation laser energy. We assume that among the three different A$_g$ modes,  A$_g^2$ in particular couples to the electronic band structure at certain resonance energies, resulting in modified Raman polarization and scattering intensity. 
The Raman polarization switching is observed for a range of sample thicknesses (here 4 to 77L). This suggests that the electronic band structure does not change drastically as thickness changes, contrary to MoS$_2$, for example \cite{splendiani2010emerging,ROSATI2024}. This might be an indication that the bulk bandgap is similar to the few-layer gap due to the extreme quasi-1d character of the material. Further studies are needed to understand this surprising observation.
Our approach, using polarization and energy-dependent Raman scattering, provides detailed insights into the anisotropic optical and vibrational characteristics of CrSBr and can be applied to different sample structures and experimental conditions. \\

\textbf{Methods}\\
\indent\textbf{Sample Fabrication}\\
Bulk CrSBr crystals were fabricated through chemical vapor transport \cite{klein2022control}. The samples were prepared through mechanical exfoliation onto SiO$_2$/Si substrates with an 85 nm SiO$_2$ layer. To probe the topology of the CrSBr flakes Oxford Instruments Cypher atomic force microscope with AC160 cantilevers was used.\\
\indent\textbf{Atomic Force Microscopy}\\
Atomic force microscopy measurements were performed at room temperature on a Cypher AFM (Asylum Research/Oxford Instruments, Wiesbaden, Germany). Height images of CrSBr flakes were obtained in AC tapping mode using the cantilever AC160TSA-R3 (300 kHz, 26 N/m, 7 nm tip radius). Images were post-processed with the in-built software features of IGOR 6.38801 (16.05.191, Asylum Research, Santa Barbara, CA, USA).\\
\indent\textbf{Photoluminescence Spectroscopy}\\
PL polarization-dependent measurements were performed using a home-built spectroscopy setup \cite{shree2021guide}. For above-band-gap excitation HeNe laser (E$_Lc$ = 1.96 eV, Thorlabs) and laser diodes (E$_Lc$ = 2.33 eV and E$_Lc$ = 1.58 eV, Thorlabs) were used. The incident light was polarized using a Glan-Laser prism and then rotated with an achromatic half-wave plate placed in front of the objective (50x, CryoGlass Optics). The sample is placed on a 3-axis nano-positioner. The objective (CryoGlass, 50x) in reflection configuration focused the laser beam on the sample surface at normal incidence. The spot size was in diameter of the order of the wavelength used, confirmed experimentally for different wavelengths by scanning the reflection signal over a metal stripe. Spectral purity of the excitation light was ensured by using corresponding MaxLine filters (Semrock). PL signal emitted by the sample was collected by the same objective and sent via free space to the spectrometer (Teledyne Princeton Instruments SpectraPro HRS-500) coupled with liquid nitrogen-cooled CCD (Teledyne Pylon 100BR excelon). The collected signal was dispersed with 150 lines/mm diffraction grating. The absence of the parasitic signal was provided by a long-pass filter (800 nm) and spatial filtering with a slit. To provide more accurate polarization-resolved dependencies a liquid crystal rotator instead of a half-wave plate was used proving the Raman mode switching effect. All measurements were conducted at ambient conditions.\\
\indent\textbf{Raman Spectroscopy}\\
Raman spectroscopy was performed in the setup described in the previous section. The same excitation sources used for PL spectroscopy were used to achieve narrow spectral lines. For the excitation in the wide wavelength range from 1.96 to 2.33 eV we utilized a continuous white light fiber laser (Fianium FIU-15, NKT Photonics) coupled with tunable high-contrast filter (LLTF Contrast VIS HP8, NKT Photonics) providing the spectral line width of $\sim$1 nm. The average power for all measurements was maintained at $\sim$100 $\mu$W. Polarization-resolved measurements were carried out with a rotated half-wave plate before the objective or liquid crystal rotator. A spectral cleaning of the laser excitation signal and blocking of the Rayleigh line of the collected signal was made with tunable short-pass and long-pass filters respectively (VersaChrome Edge Filters, Semrock). To achieve higher spectral resolution of narrow Raman lines we used 1200 lines/mm diffraction grating. All measurements were conducted at ambient conditions.\\
\indent\textbf{Photoluminescence Excitation Spectroscopy}\\
The same continuous white light fiber laser with a tunable filter was used as an excitation source for PLE measurements. The average incident power of 100 $\mu$W was maintained at each wavelength from 1.46 to 2.76 eV with a gradient ND filter wheel and measured before the objective. The polarization of excitation light was set along a or b crystallographic axis of specific CrSBr flake.\\

\textbf{Data availability}\\
The data that support the findings of this study are available from the corresponding
authors upon request.\\

 \textbf{Acknowledgements}\\
 This work is financially supported by supported by ERC-CZ program (project LL2101) from Ministry of Education Youth and Sports (MEYS) and by the project Advanced Functional Nanorobots (reg. No. CZ.02.1.01/0.0/0.0/15$\_$003/0000444 financed by the ERDF). \\

\textbf{Author contributions} 
K.M. and Z.S. grew the bulk CrSBr crystals. L.H. and P.M. fabricated few-layer CrSBr samples for optical spectroscopy and with D.I.M. performed optical spectroscopy measurements. P.M., K.H., and R.v.K. performed and interpreted AFM measurements. P.M., D.I.M., L.H and B.U. analyzed the optical spectra. M.M.G. and I.G. contributed to the theoretical explanation of observed phenomena. P.M., D.I.M., L.H., L.K., S.S., J.K., R.v.K., M.M.G. and B.U. discussed the results. B.U. suggested the experiments and supervised the project. P.M., D.I.M., and B.U. wrote the manuscript with input from all the authors.\\

\textbf{Competing interests}: The authors declare no competing interests.\\


\newpage

   

\renewcommand{\thefigure}{S\arabic{figure}}

\setcounter{figure}{0}

 \begin{figure*}
\includegraphics[width=1\linewidth]{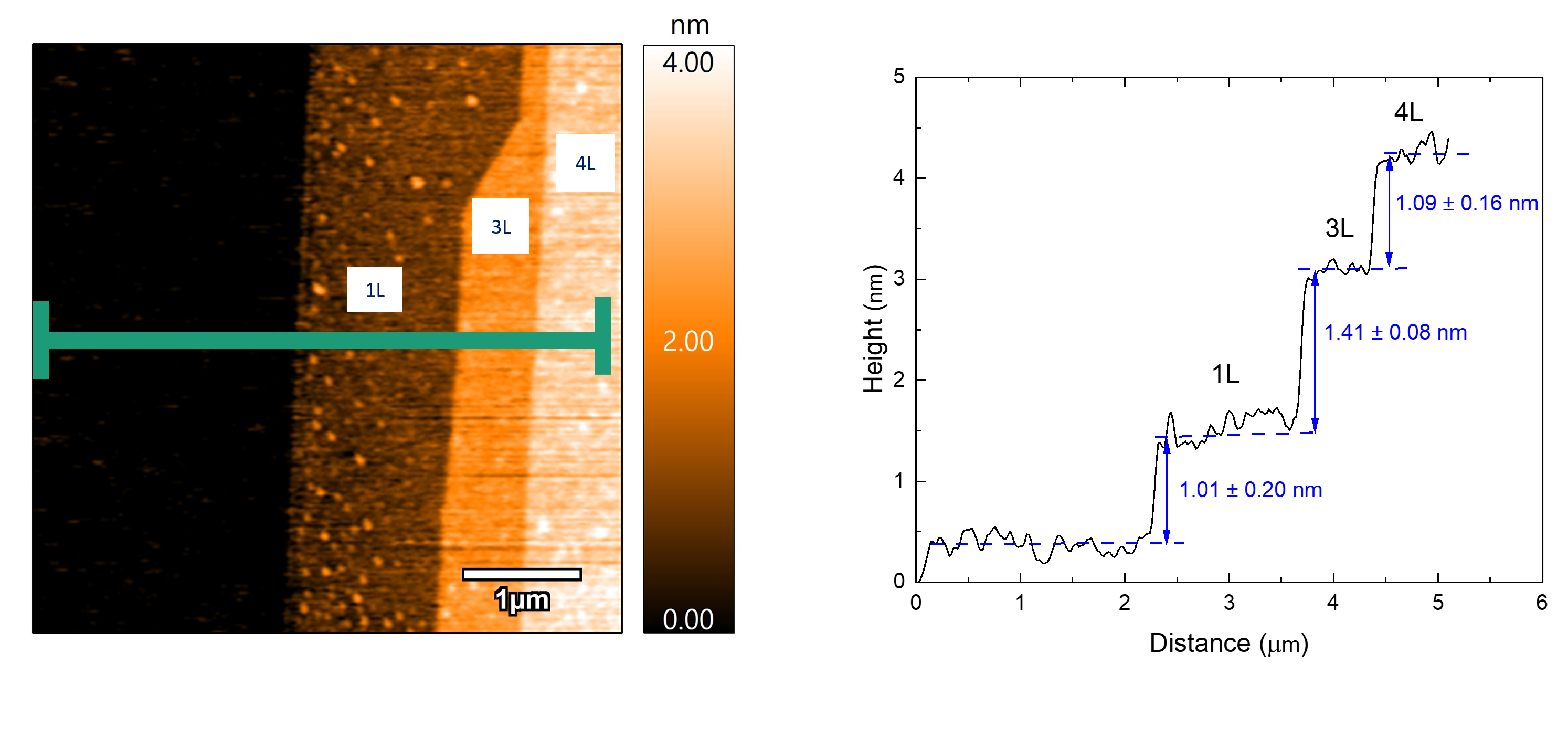}
\caption{\label{fig:figS1}\textbf{Atomic force microscopy (AFM).} On the left the AFM image with a green line indicating the position of the thickness profile.The corresponding thickness profile on the right, where the monolayer thickness is measured as 1.01 nm.}
\end{figure*}

\begin{figure*}
\includegraphics[width=1\linewidth]{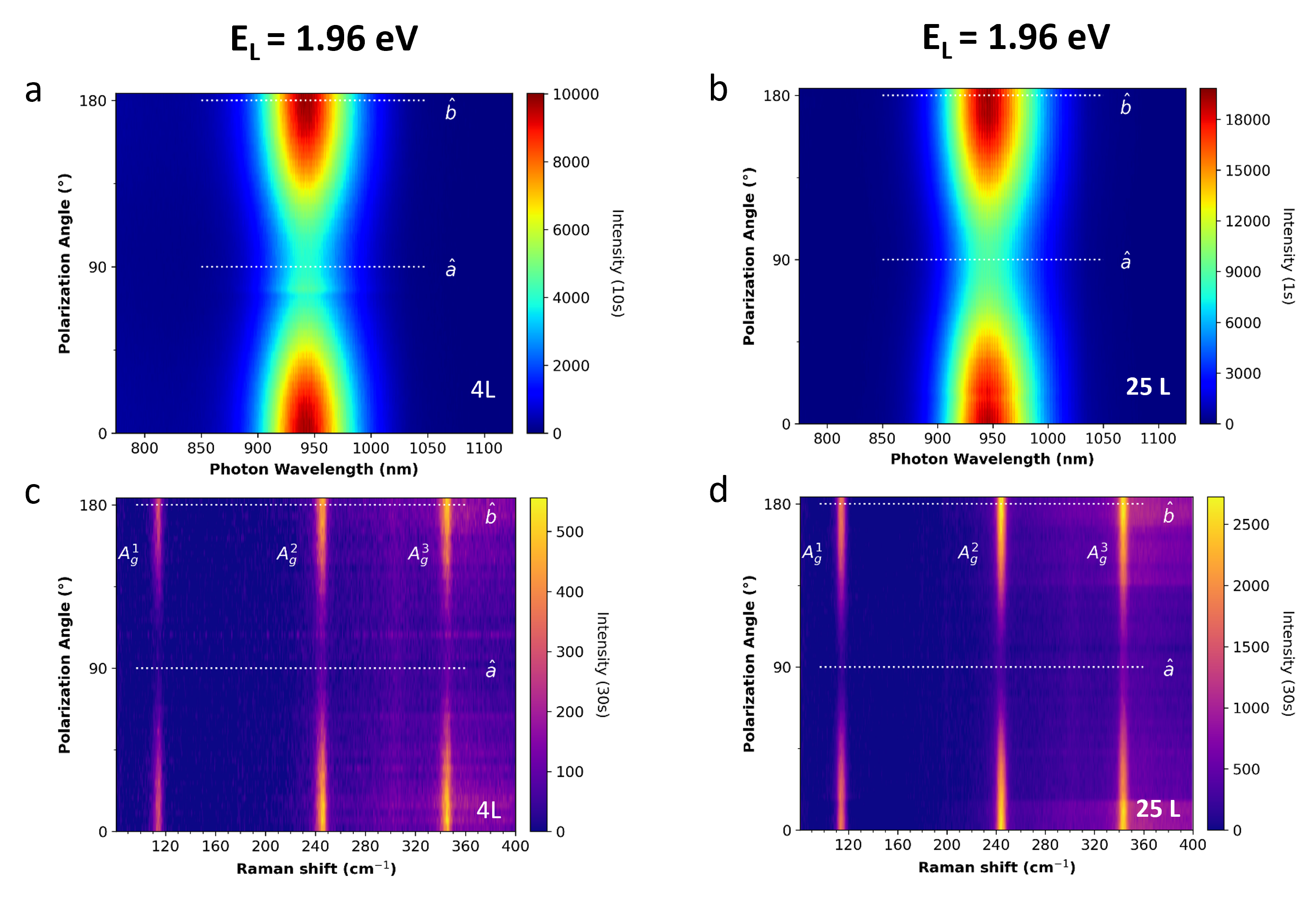}
\caption{\label{fig:figS2}\textbf{Polarization-resolved measurements.} The top panel shows the polarization-resolved photoluminescence, and the bottom panel shows the polarization-resolved Raman data for two different thicknesses: 4L (left panel) and 25L (right panel), both with a laser energy of 1.96 eV.}
\end{figure*}

\begin{figure*}
\includegraphics[width=1\linewidth]{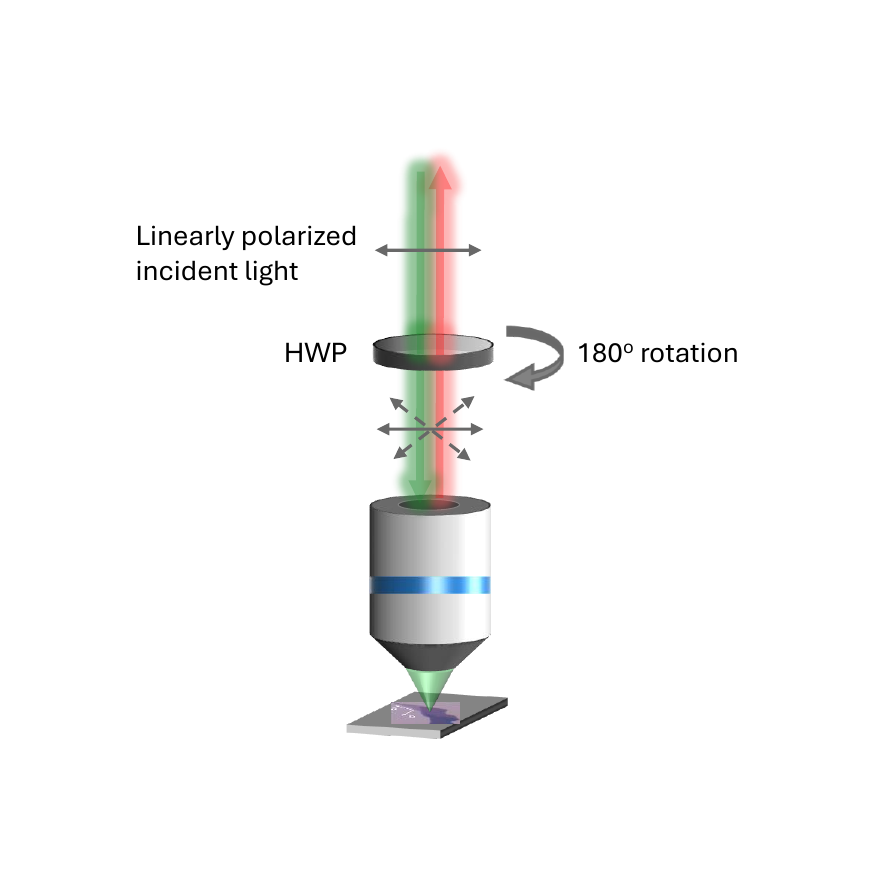}
\caption{\label{fig:figS3}\textbf{Schematic of the polarization optics for the Raman spectroscopy measurements using a half-wave-plate (HWP).}}
\end{figure*}

\begin{figure*}
\includegraphics[width=1\linewidth]{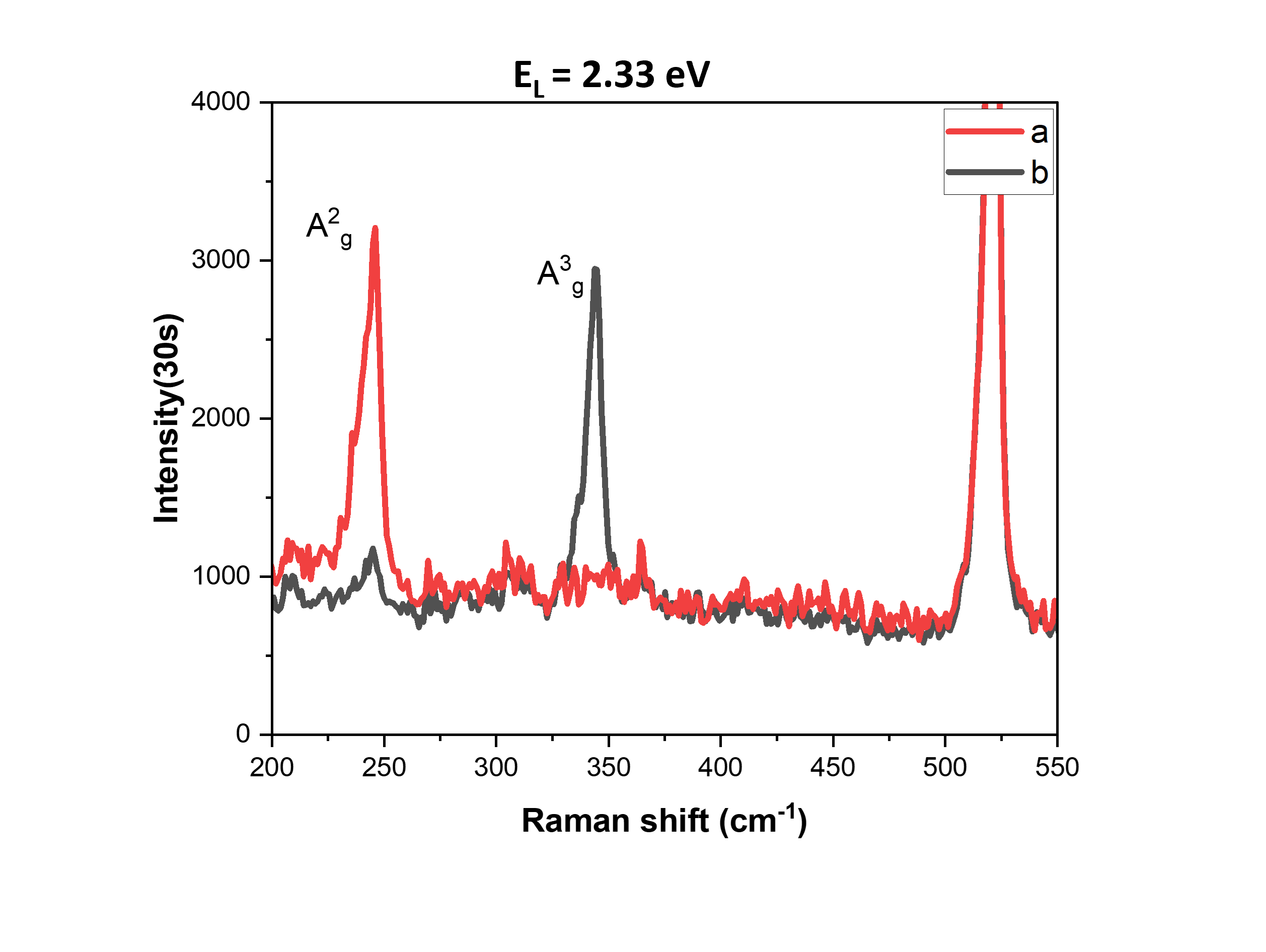}
\caption{\label{fig:figS4}\textbf{Raman spectroscopy for Laser energie 2.33 eV.} Shown here is a typical Raman spectrum obtained using the NKT Laser (see Methods), which allows for the extraction of polarization dependence.}
\end{figure*}

\begin{figure*}
\includegraphics[width=1\linewidth]{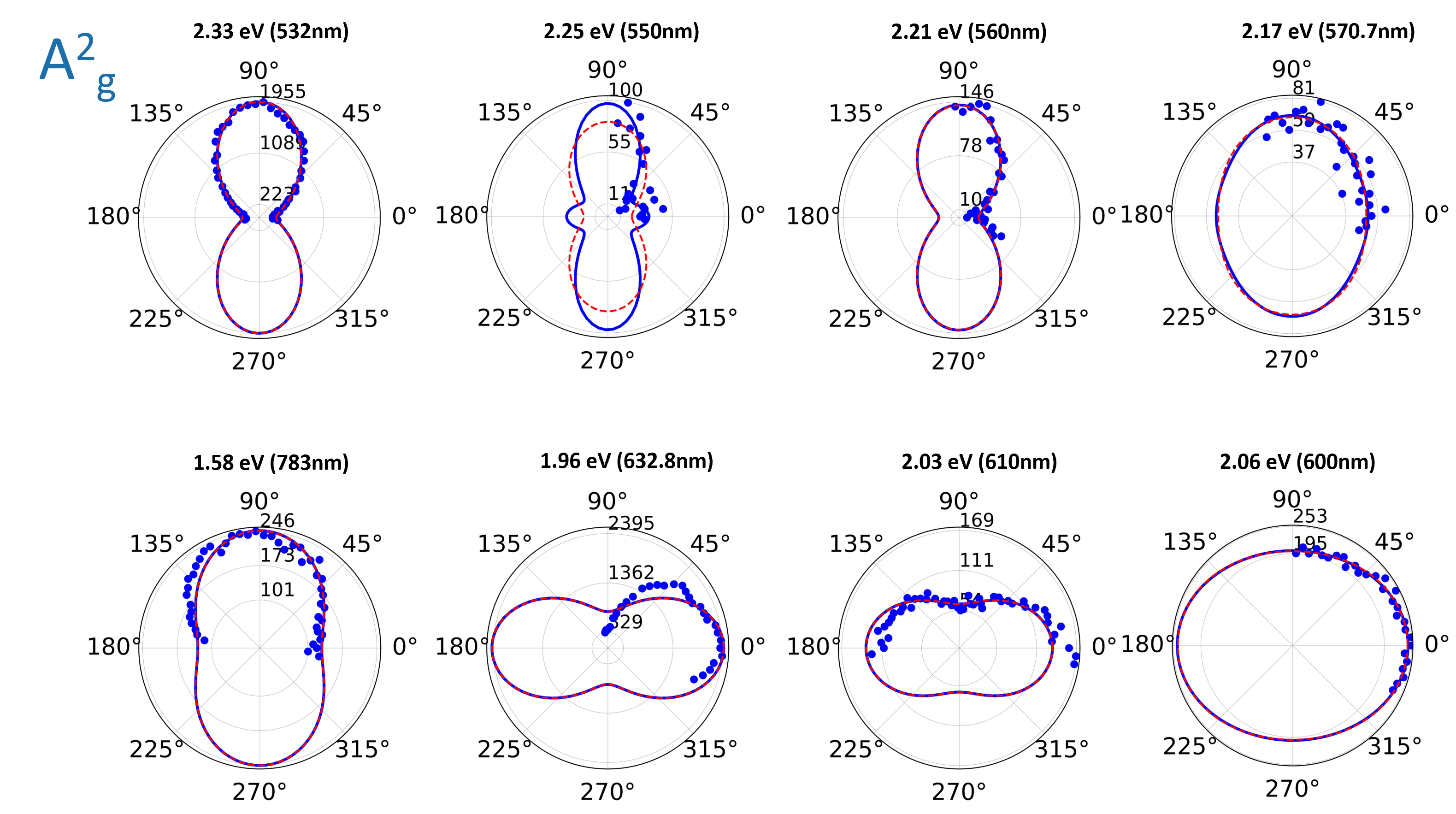}
\caption{\label{fig:figS5} \textbf{Polarization-resolved Raman.} Polarization-resolved Raman data for the A$_g^2$ mode for different excitation laser wavelengths for the 6L sample. The red dotted curve is fitted with $\phi_{ab} = 0$ and the blue curve is fitted with $\phi_{ab} \neq 0$.}

\end{figure*}

\begin{figure*}
\includegraphics[width=1\linewidth]{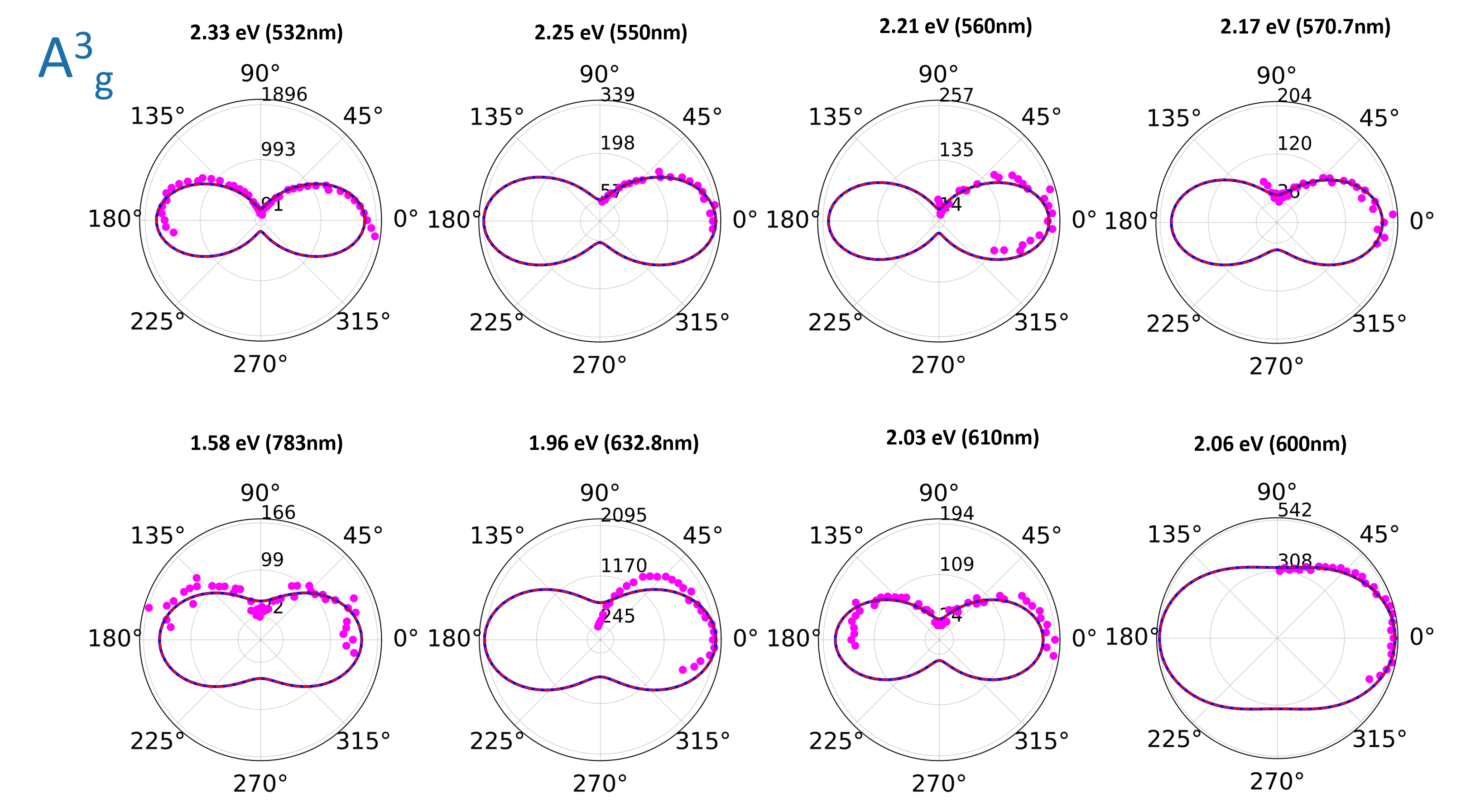}
\caption{\label{fig:figS6} \textbf{Polarization-resolved Raman.} Polarization-resolved Raman data for the A$_g^3$ mode for different excitation laser wavelengths for the 6L sample. The red dotted curve is fitted with $\phi_{ab} = 0$ and the magenta curve is fitted with $\phi_{ab} \neq 0$.}

\end{figure*}

\begin{figure*}
\includegraphics[width=1\linewidth]{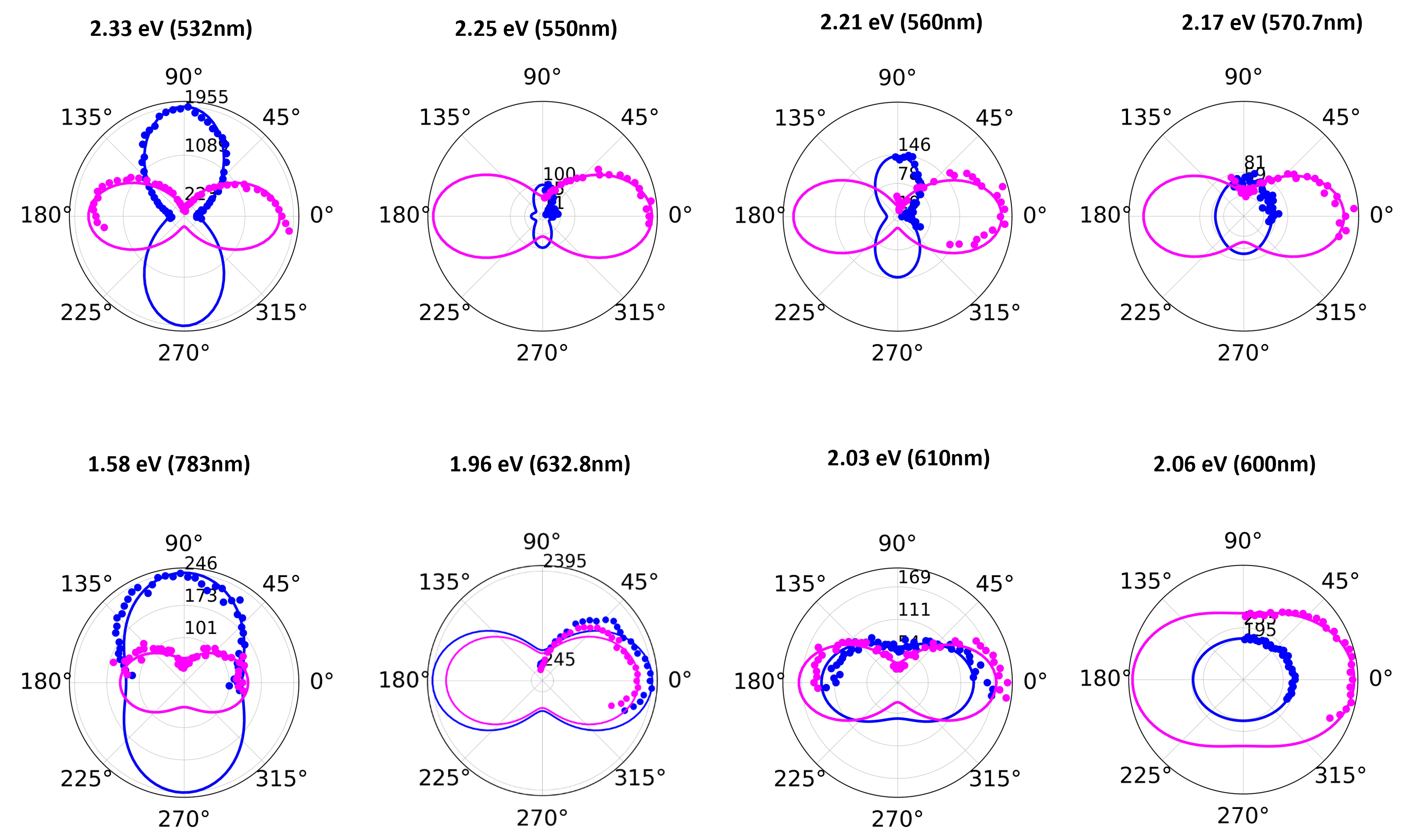}
\caption{\label{fig:figS7} \textbf{Polarization-resolved Raman.}Polarization-resolved Raman modes for different excitation laser wavelength, the intensities within each panel can be directly compared for 6L sample.A$_g^2$ and A$_g^3$ are represented with blue and magenta respectively.} 
\end{figure*}

 \begin{figure*}
\includegraphics[width=0.8\linewidth]{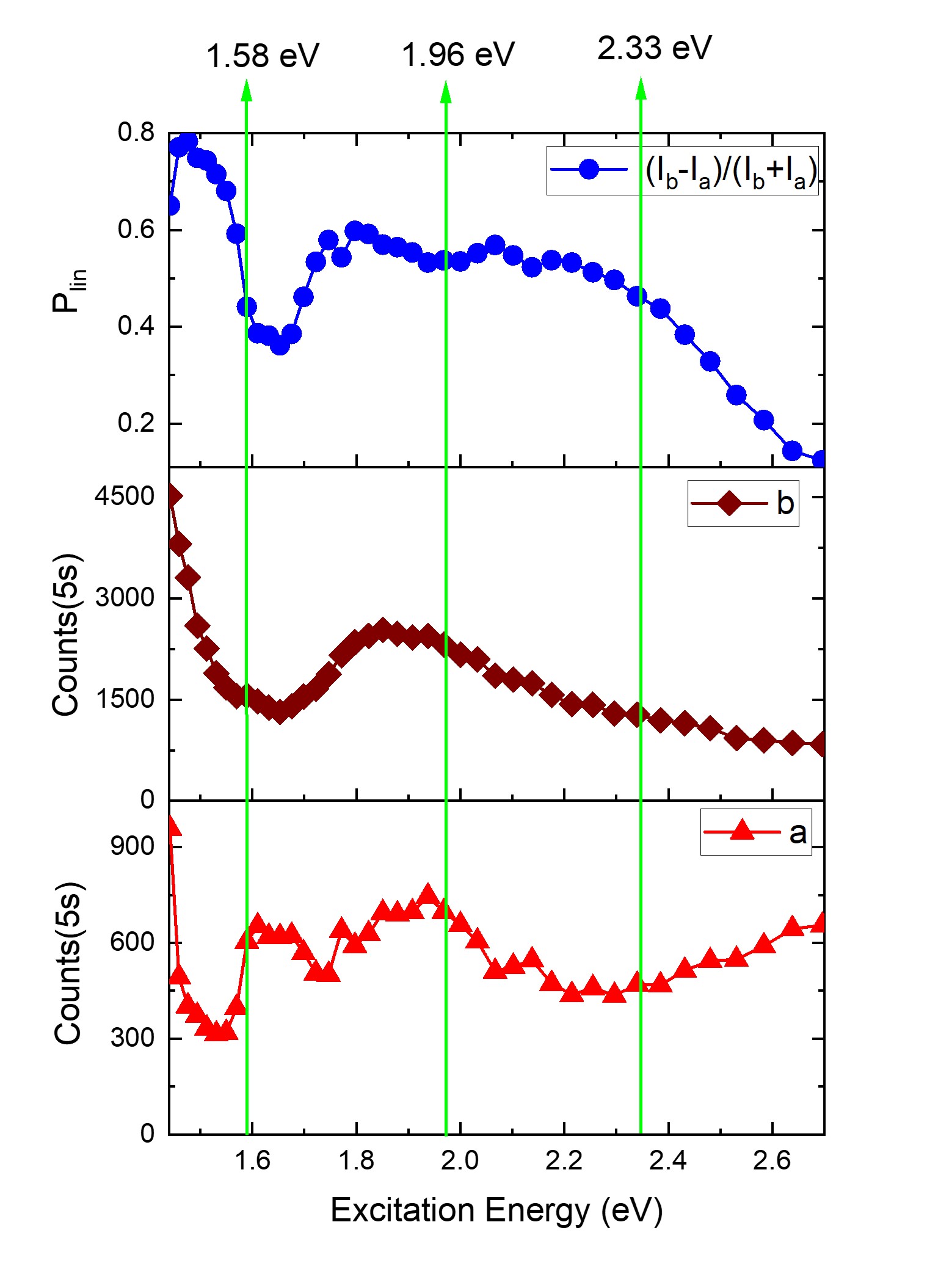}
\caption{\label{fig:figS8}  \textbf{Photoluminescence Excitation (PLE) Spectra.} This figure presents the PLE data for the 4L sample measured along both crystal axes $b$ and $a$ (middle and bottom panel respectively). The distinct differences between the two crystallographic axes are clearly evident in linear polarization degree \(P_{lin} = (I_b - I_a)/(I_b + I_a)\) (top panel), highlighting the anisotropic behavior of the sample.}
\end{figure*}

\begin{figure*}
\includegraphics[width=1\linewidth]{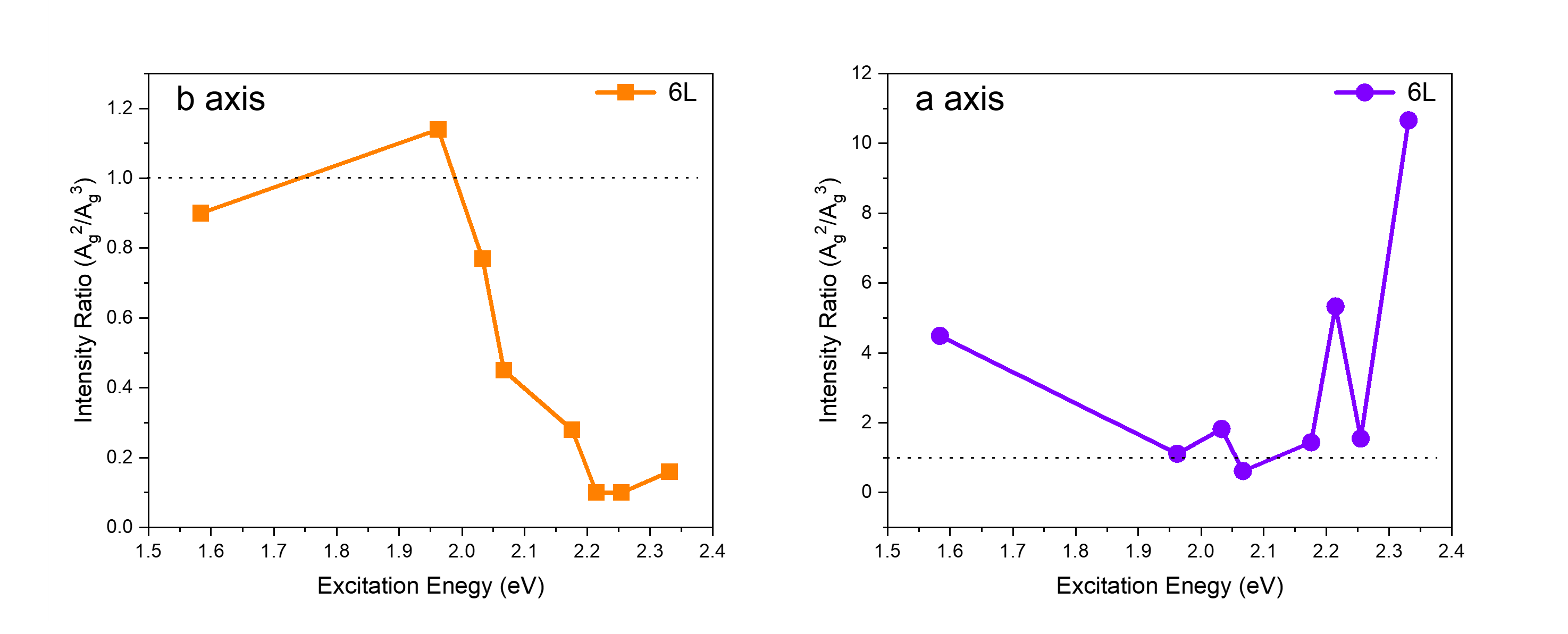}
\caption{\label{fig:figS9}\textbf{Intensity ratios of two Raman modes.} The intensity ratios A$_g^2$ and A$_g^3$ are shown for laser polarization along the crystal axes, with the b-axis (left) and the a-axis (right).}
\end{figure*}

\begin{figure*}
\includegraphics[width=1\linewidth]{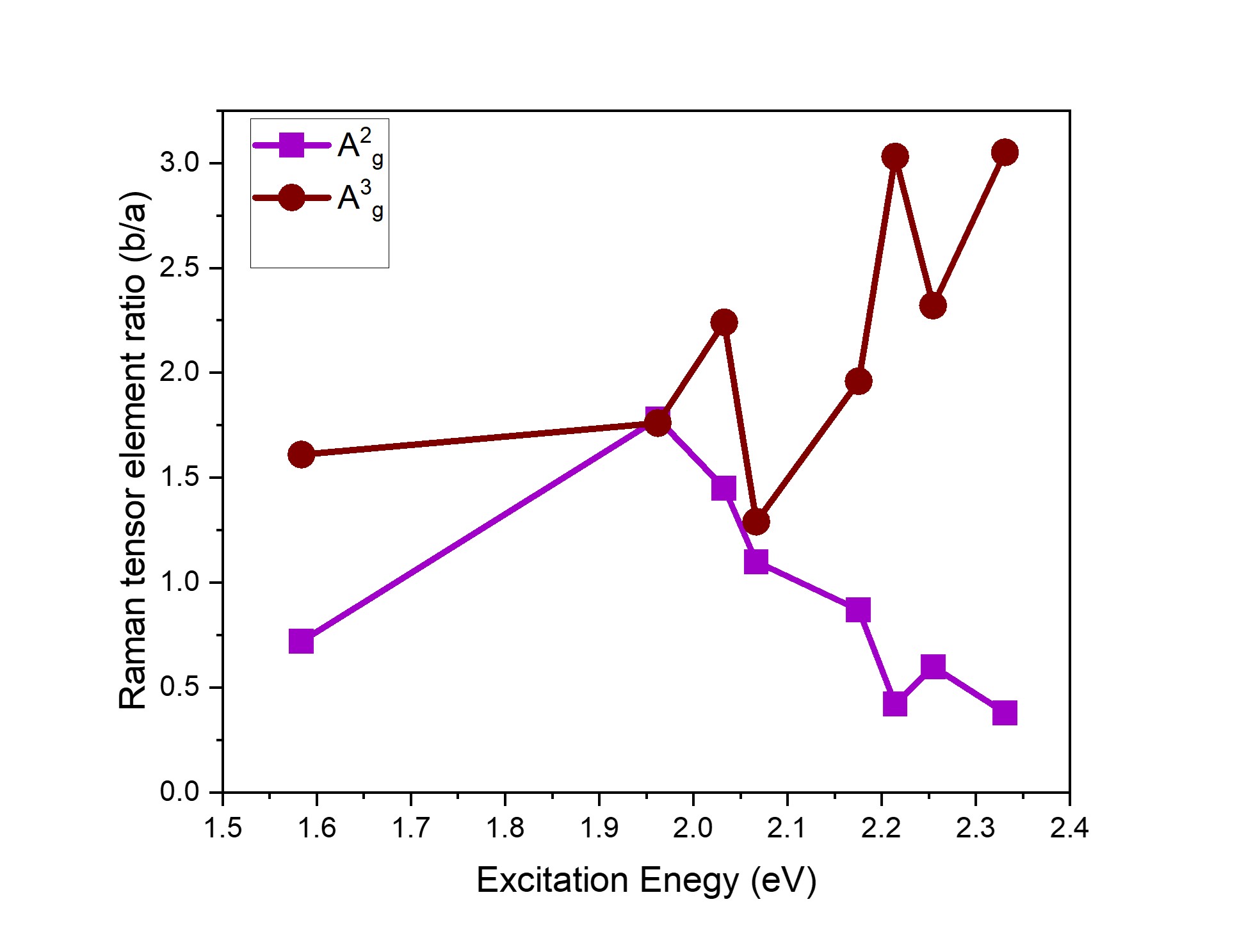}
\caption{\label{fig:figS10} \textbf{Ratios of two Raman tensor elements.} The ratios are shown for the A$_g^2$ and A$_g^3$ modes at different laser excitation energies.}
\end{figure*}

 \begin{figure*}
\includegraphics[width=1\linewidth]{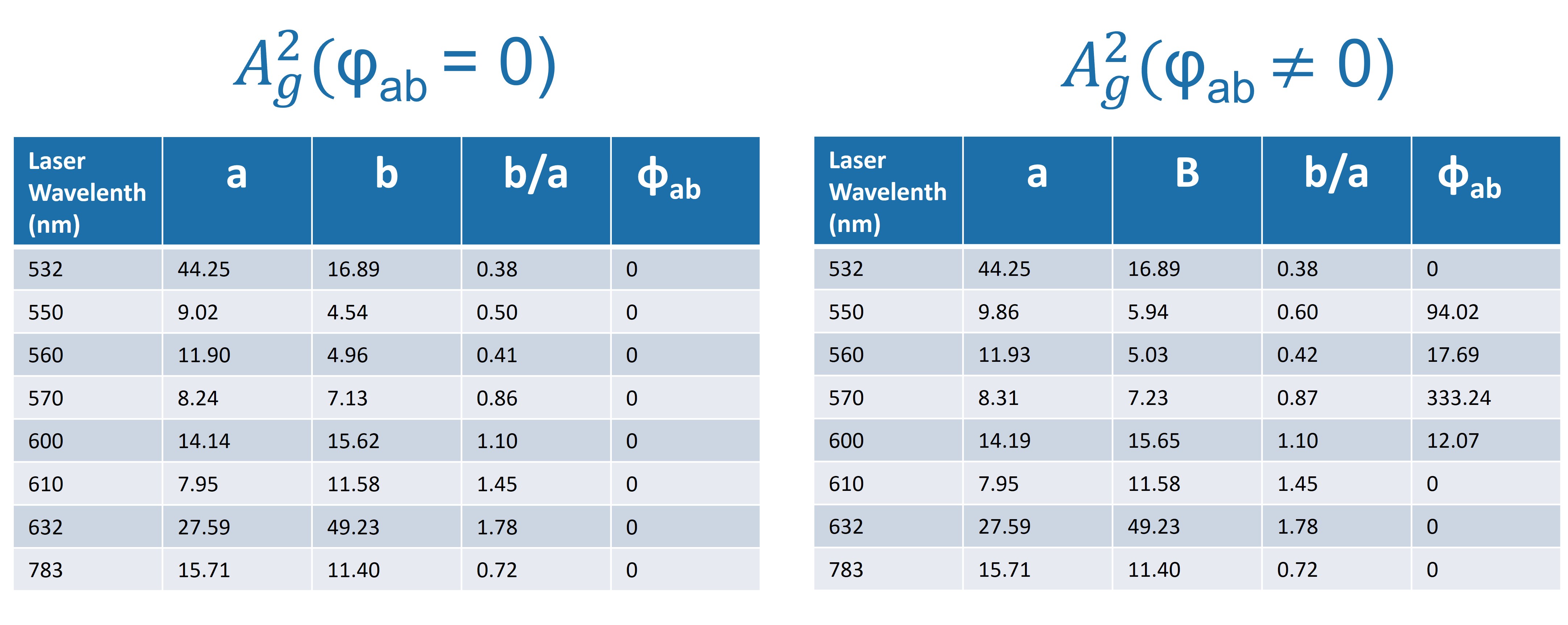}
\caption{\label{fig:figS11} \textbf{Raman Intensity Angular Fitting for A$_g^2$ .} The Raman intensity as a function of angle is fitted using Equation 2 from the main text. The ratios of two Raman tensor elements for the A$_g^2$ mode are shown for both fixed phase $\phi_{ab}=0$ and when $\phi_{ab}$ is allowed to vary as a free fitting parameter.}
\end{figure*}

 \begin{figure*}
\includegraphics[width=1\linewidth]{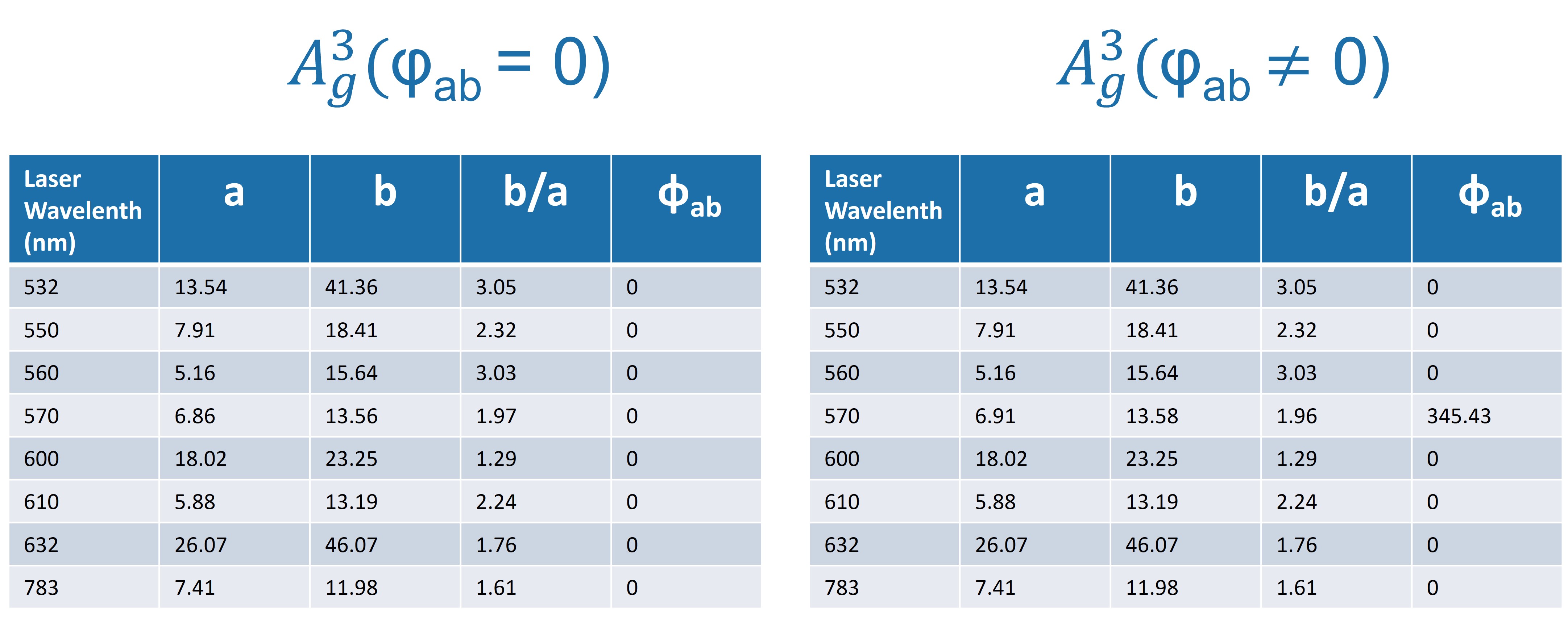}
\caption{\label{fig:figS12}\textbf{Raman Intensity Angular Fitting for A$_g^3$ .} The Raman intensity as a function of angle is fitted using Equation 2 from the main text. The ratios of two Raman tensor elements for the A$_g^3$ mode are shown for both fixed phase $\phi_{ab}=0$ and when $\phi_{ab}$ is allowed to vary as a free fitting parameter.}
\end{figure*}

\end{document}